\documentclass[doublecol]{epl2} 

\usepackage{graphicx}
\usepackage{dcolumn}
\usepackage{bm}
\usepackage{mciteplus}
\usepackage{braket}
\usepackage{upgreek}
\usepackage{amsmath}
\usepackage{amssymb}

\usepackage[svgnames]{xcolor}

\renewcommand{\section}[1]{\textit{#1}.---}
\usepackage[colorlinks=true,linkcolor=DarkBlue,urlcolor=DarkBlue,citecolor=FireBrick]{hyperref}

\title{Markov property of Lagrangian turbulence}
\shorttitle{Markov property of Lagrangian turbulence} 

\author{A.~Fuchs\inst{1} \and M.~Obligado\inst{2} \and M.~Bourgoin\inst{3} \and M.~Gibert\inst{4} \and P.D.~Mininni\inst{5} \and J.~Peinke\inst{1}}

\institute{                    
	\inst{1} Institute of Physics and ForWind, University of Oldenburg, Küpkersweg 70, 26129 Oldenburg, Germany\\
	\inst{2} Univ. Grenoble Alpes, CNRS, Grenoble INP*, LEGI, 38000 Grenoble, France\\
	\inst{3} Laboratoire de Physique de l'\'Ecole Normale Sup\'erieure de Lyon, CNRS \& Universit\'e de Lyon, 46 all\'ee d’Italie, F-69364 Lyon Cedex 07, France\\
	\inst{4} Univ. Grenoble Alpes, CNRS, Grenoble INP, Institut Néel, 38000 Grenoble, France\\
	\inst{5} Universidad de Buenos Aires, Facultad de Ciencias Exactas y Naturales, Departamento de F\'\i sica, \& IFIBA, CONICET, Ciudad Universitaria, Buenos Aires 1428, Argentina
}

\abstract{
	Based on direct numerical simulations with point-like inertial particles, with Stokes numbers, $\textrm{St}=0, 0.5$, $3$, and $6$, transported by homogeneous and isotropic turbulent flows, we present in this letter for the first time evidence for the existence of Markov property in Lagrangian turbulence.
	We show that the Markov property is valid for a finite step size larger than a Stokes number-dependent Einstein-Markov coherence time scale. 
	This enables the description of multi-scale statistics of Lagrangian particles by Fokker-Planck equations, which can be embedded in an interdisciplinary approach linking the statistical description of turbulence with fluctuation theorems of non-equilibrium stochastic thermodynamics and local flow structures. 
	The formalism allows estimation of the stochastic thermodynamics entropy exchange associated with the particles’ Lagrangian trajectories. 
	Entropy consuming trajectories of the particles are related to specific evolution of velocity increments through scales and may be seen as intermittent structures.
    Statistical features of Lagrangian paths and entropy values are thus fixed by the fluctuation theorems.
}

\begin{document}
	
	\maketitle
	
	\section{Introduction}
	The physics of particles submerged in fluids has played a central role in the development of statistical mechanics, and in our current understanding of out-of-equilibrium systems. 
	Inertial particles are inclusions in the flow which are denser or lighter than the fluid, and have a size smaller or larger than the smallest relevant flow scale (the scale of the smallest eddies, also called the flow dissipative scale). 
	Such particles are carried by the fluid, but they also have their own inertia, and therefore the fluid and the particles have different dynamics. 
	In the limit of point-wise particles with negligible inertia, the particles become Lagrangian tracers, which perfectly follow the fluid elements. 
	In all these two-phase systems the mechanisms that explain how turbulence affects the motion of the particles are not completely clear. 
	As an example, turbulence can both enhance or hinder the settling velocity of inertial particles \cite{Falkinhoff_2020}. 
	For heavy particles, an initially homogeneous distribution of particles may, after interacting with a turbulent flow, regroup into clusters forming dense areas and voids, in a phenomenon called preferential concentration where turbulence somehow unmixes the particles \cite{Mora_2021}.
    Thus, the velocity and statistics of tracers and inertial particles is also different: while tracers sample the flow homogeneously, inertial particles only see specific regions of the flow, with either low acceleration or vorticity, and their velocity can differ significantly from the fluid velocity.
	It is only in the last decades when sufficient time and spatial resolution have been achieved in experiments and numerical studies to allow the analysis of these phenomena. 
	Frequently new numerical and experimental data has been in contradiction with theoretical models, and previous knowledge on fluid-particle interactions had to be reconsidered even in simplified cases \cite{Toschi_2009}. 
	Furthermore, there are many open questions concerning inhomogeneous flows \cite{Stelzenmuller_2017}, finite-size \cite{Fiabane_2012,Cisse_2015} and non-spherical particles~\cite{Borgnino_2019}, among others. 
	More recently, colloidal particles were used in the first experiments to verify fluctuation relations such as the Jarzynski equality \cite{Toyabe_2010}, which links the statistics of fluctuating quantities in a non-equilibrium process with equilibrium quantities. 
	Colloidal particles were also used to verify the thermodynamic cost of information processing \cite{B_rut_2012} proposed by Landauer. 
	Fluctuation theorems for more general problems, in which a physical system is coupled with (and driven by) another out-of-equilibrium system, would open potential applications in other complex coupled systems as in soft and active matter. 
	But what fundamental statistical relations are satisfied by particles that interact with a complex and out-of-equilibrium turbulent flow? 
	Brownian motion set the path for the study of diffusion and random processes.
	It becomes more complicated if the particle motion is driven by turbulence; even a simple point-wise passive particle provides a challenge in the modeling of many aspects of its dynamics \cite{Falkovich_2002, Bourgoin_2014}.\\

	\section{Friedrich-Peinke approach to turbulence}
	In this letter we show that 
	the Markov property is valid for the dynamics of dense sub-Kolmogorov particles coupled to a turbulent velocity field of three-dimensional homogeneous and isotropic turbulence (HIT) at inertial range time scales. 	
	To this end we use pseudo-spectral direct numerical simulations (DNSs) solving the incompressible Navier-Stokes equation with a simple point particle model \cite{Mininni_2011}.
	To characterize HIT it is customary to study the statistics of Eulerian velocity increments \mbox{$u_r = \lbrack \textbf{u}(\textbf{x}+\textbf{r})-\textbf{u}(\textbf{x}) \rbrack \cdot \textbf{r}/\vert \textbf{r} \vert$} for scales $r=|\textbf{r}|$. 
	However, two-point statistics of velocity increments do not fully characterize small-scale turbulence \cite{renner2002universality}, and many attempts at dealing with multi-point statistics have been considered. 
	One way to do this is to use the Friedrich-Peinke approach \cite{friedrich1997description}, in which the stochastic dynamics of velocity increments $u_r$ are considered as they go through the cascade from large to small scales $r$. 
	A central assumption is that the evolution of the stochastic variable $u_r$ possesses a Markov process ``evolving'' in $r$. Previous studies showed that $u_r$ can be considered as Markovian to a reasonable approximation~\cite{friedrich1997description, Marcq_1998, renner2001}, at least down to a scale close to the Taylor scale \cite{Lueck2006markov, renner2001}. 
	Furthermore, $u_r$ satisfies a diffusion process \cite{GardinerBook}
	\begin{eqnarray}
		\label{eq:Langevin}
		-\partial _r  u_{r} = D^{(1)} (u_{r},r) + [D^{(2)} (u_{r},r)]^{1/2} \Gamma(r),
	\end{eqnarray}
	where a linear dependence on the value of the increment for the drift coefficient $D^{(1)}(u_r, r)$, and a quadratic dependence for the diffusion coefficient $D^{(2)}(u_r, r)$, was found \cite{renner2001, renner2002universality, Nawroth2007, Reinke_2018, Fuchs_2020}.
	Further details on the noise term $\Gamma(r)$ and the empirical estimation of the drift and diffusion coefficients $D^{(1,2)}(u_r, r)$ are given in \cite{SM} (Appendix A) .

	Up to now, this approach has been applied to an Eulerian description of turbulent flows.
	Based on the general interest in the relation between Eulerian and Lagrangian properties of turbulence the question of the application of this approach and, in particular, the verification of the necessary condition of Markov property
	for Lagrangian turbulence and inertial particles arises.\\
	
	\section{DNS of inertial particles in turbulence}
	To this end we use DNSs of forced HIT.
	A large-scale external mechanical forcing is given by a superposition of modes with slowly evolving random phases, following standard practices for its temporal integration and de-aliasing procedures. 
	An adequate spatial resolution of the smallest scales, i.e., $\kappa\eta \gtrsim 1$ is chosen  \cite{pope2001turbulent}. 
	Here, $\eta$ is the Kolmogorov or dissipation length scale, $\eta = (\nu^3/\varepsilon)^{1/4}$ (where $\varepsilon$ is the kinetic energy dissipation rate, and $\nu$ the kinematic viscosity of the fluid), and $\kappa=N_\textrm{DNS}/3$ is the maximum resolved wavenumber in Fourier space (with $N_\textrm{DNS}=512$ the linear spatial resolution in each direction). 
	The DNSs are characterized by a Reynolds number based on the Taylor microscale of $Re_\lambda \approx 240$ (see \cite{Mora_2021} for details). 
	Inertial particles were modeled using the Maxey-Riley-Gatignol equation in the limit of point heavy particles, which for a particle with velocity $\mathbf{v}$ in the position $\mathbf{x_p}$ submerged in a flow with velocity $\mathbf{u}$, reads 
	\begin{equation}
		\dot{\mathbf{v}}(t) = [\mathbf{u}(\mathbf{x_p},t)-\mathbf{v}(t)]/{\tau_p},
		\label{eq:inertial}
	\end{equation}
	with $\tau_p$ the particle Stokes time. 
	For tracers, $\tau_p=0$.
	There is no particle-particle or particle-fluid interaction (i.e., we use a one-way coupling approximation). 
	The distinction between particles is quantified by the Stokes number $\textrm{St} = \tau_p/\tau_\eta$ (the ratio of the particle relaxation time to the Kolmogorov time, defined as $\tau_\eta = [\nu/\epsilon]^{1/2}$).\\

	\section{Cascade trajectories}
	To study the Markov property of particles trajectories, we study $\mathbf{v}(t)$ (the Lagrangian velocity of the particle at time $t$) for $\textrm{St}>0$, and for Lagrangian tracers with $\textrm{St}=0$ the analysis is performed using $\mathbf{v}(t)=\mathbf{u}(\mathbf{x_p},t)$ (the velocity of the fluid element where the tracer is located).
	Building now velocity increments in time 
	instead of the commonly used spatial scales $r$ ($\mathbf{e_i}$ corresponds to the unitary vector in the direction $x$, $y$, or $z$ specified in the figures)
	\begin{eqnarray}
		u_\tau = [\mathbf{v}(t+\tau)-\mathbf{v}(t)]\cdot \mathbf{e_i},
	\end{eqnarray}
	we define for every particle trajectory a ``cascade trajectory" (or ``sequence" of velocity increments at decreasing scales) $\left[u(\cdot)\right]=\{u_T,\dots,u_{\uptau_f} \}$ for different time-separations or time scales $\tau$, from the initial $T$ to the final time scale $\uptau_f$ with $T>\uptau_f$. 
	The notation $\left[u(\cdot) \right]$ indicates the entire path through the hierarchy of time scales from large to small scales
	instead of a distinct value $u_\tau$; $T \approx 120 \tau_\eta$ and $\uptau_f \approx 0.2 \tau_\eta$ were chosen as references.
	The advantage of a Lagrangian perspective compared to a spatial (Eulerian) perspective is that Lagrangian particle trajectories have a clear physical meaning: each particle follows a path according to the fluid dynamics and its own inertia. 
	In other words, the time increments have a direct interpretation as values corresponding to changes in the velocity along trajectories as the system evolves. 
	In the Eulerian case, the spatial distances (or the ``evolution'' in $r$) have an indirect interpretation in the context of a Fokker-Planck formulation of the dynamics.\\

	\section{Results} 
\begin{figure}[ht]
		\includegraphics[width=0.23\textwidth]{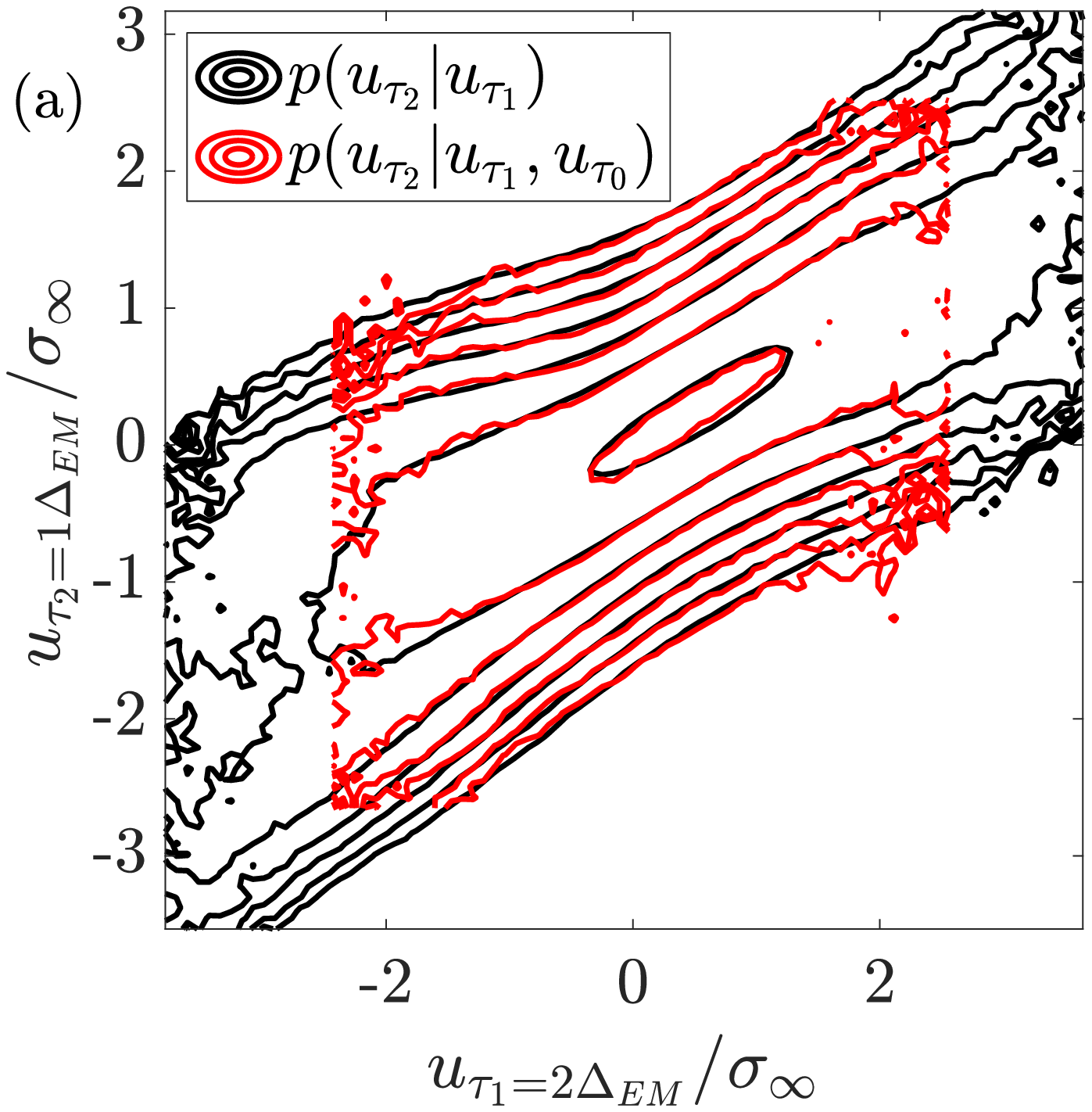}
		\includegraphics[width=0.23\textwidth]{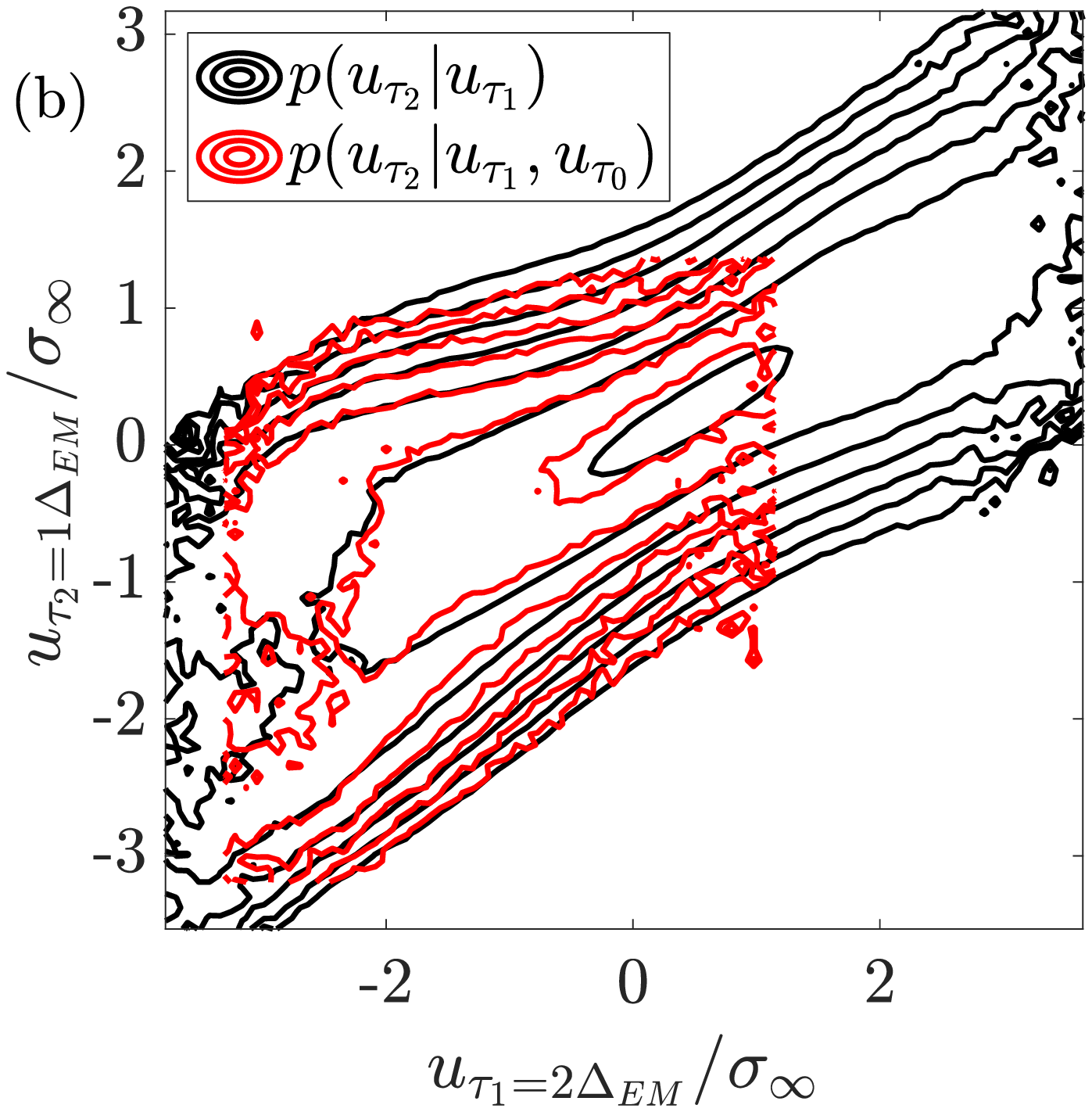}
		\caption{Visualization of Markov properties at $\textrm{St}=3$. Contour plots showing single conditioned (black solid lines) and double conditioned PDFs (red solid lines) of velocity increments for three different time scales $\tau_0>\tau_1>\tau_2$, each separated by $\Delta \tau = \Delta_{EM}$, which is the Einstein-Markov coherence time scale for $\textrm{St}=3$. The conditioned value for the large increment was chosen here as $u_{\tau_{0}}=0$ (a) and $u_{\tau_{0}}=-2$ (b). Note the choice of $u_{\tau_{0}}$ changes the amount of events as well as the center of the double conditioned PDF. We use the normalization of the increments introduced in \cite{renner2001} with \mbox{$\sigma_{\infty} = \sqrt{2}\sigma$}, where $\sigma$ is the data standard deviation.}
		\label{fig2}
	\end{figure}

\begin{figure}[ht]
		\includegraphics[width=0.48\textwidth]{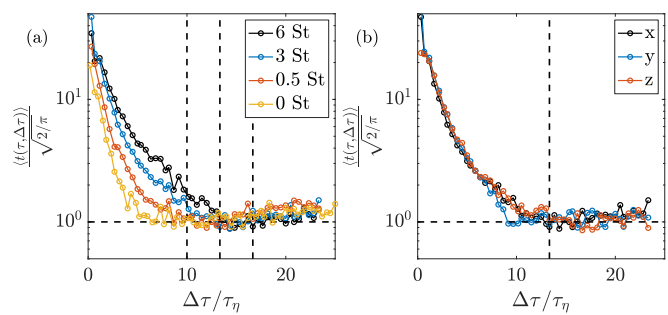}
		\caption{Wilcoxon test for the particles trajectories: for the $x$ component of the particles velocity for all Stokes numbers studied here as function of $\Delta\tau$ (a). Same test but for all three velocity components at $\textrm{St}=3$ (b). Note, that the normalized expectation value $t(\tau,\Delta \tau)$  is supposed to be close to 1 if the Markovian assumption is valid (see  \cite{renner2001,Lueck2006markov} and \cite{SM} (Appendix G) for details and the description of essential notions).} 
		\label{fig1}
	\end{figure}

	To verify the existence of the Markov property
	for Lagrangian turbulence, we integrate three sets of $1.3 \times 10^5$ inertial particles, respectively with \mbox{$\textrm{St}=0, 0.5$, $3$, and $6$}.
	In Fig.~\ref{fig2} a qualitative validation of the Markov property based on the alignment of the single conditioned $p\left(u_{\tau_2}|u_{\tau_1}\right)$ and double conditioned $p\left(u_{\tau_2}|u_{\tau_1}, u_{\tau_0}\right)$ probability density functions (PDFs) of datasets of  increments for a chosen set of three different time scales $\tau_0>\tau_1>\tau_2$ is shown as contour plots. 
	Each scale is separated by $\Delta \tau \approx 13 \tau_\eta$, which is equal to the Einstein-Markov coherence time scale $\Delta_{EM}$ for $\textrm{St}=3$.
	As described below and in the discussion accompanying Fig.~\ref{fig1}, $\Delta_{EM}$ is the smallest time-scale for which the Markovian assumption can be considered valid.
	For Markovian processes the relation \mbox{$p\left(u_{\tau_2}|u_{\tau_1}\right) = p\left(u_{\tau_2}|...,u_{\tau_1}, u_{\tau_0}\right)$} holds. For finite datasets,
	\begin{eqnarray}
		p\left(u_{\tau_2}|u_{\tau_1}\right) = p\left(u_{\tau_2}|u_{\tau_1}, u_{\tau_0}\right) ,
		\label{eq:markov}
	\end{eqnarray}
	is commonly assumed to be a sufficient condition.
	The close alignment between the PDFs (black and red solid lines in Fig.~\ref{fig2}) confirms the validity of the Markov property.
	Note that while Fig.~\ref{fig2} shows results for $\textrm{St}=3$, all our datasets give similar results. 
	
	In Fig.~\ref{fig1} the Markovian approximation is checked systematically for different values of $\Delta \tau$, using the Wilcoxon test, which is a quantitative and parameter-free 
	test that determines the Einstein-Markov coherence time scale $\Delta_{EM}$ (see  \cite{renner2001,Lueck2006markov} and \cite{SM}, Appendix G, for details and the description of essential notions).
	This test is a reliable procedure to validate Eq.~(\ref{eq:markov}) \cite{renner2001,Lueck2006markov}.
	Based on this analysis, the Markovian approximation is valid for time-scales larger than or equal to this $\textrm{St}$-dependent critical time separation \mbox{$\Delta_{EM}\approx 7, 10, 13$ and $16\tau_\eta$} for \mbox{$\textrm{St}=0, 0.5$, $3$, and $6$ } respectively (see Fig.~\ref{fig1}(a)).
	Accordingly, the complexity of the dynamics of inertial particles in turbulent flows, expressed by the evolution of the stochastic variables $u_\tau$ at decreasing time scales, can be treated as a Markov process, with an Einstein-Markov coherence time scale that increases with $\textrm{St}$.
	Our analysis presented in \cite{SM} \mbox{(Appendix C and D)} demonstrates that on the scale that the Markov property holds the non-Gaussian behavior is still present and that our results are well linked to the range containing Lagrangian small-scale intermittency. 
	Furthermore, we see that $\Delta_{EM}$ is the same (at fixed $\textrm{St}$) for all velocity components (see Fig.~\ref{fig1}(b)).
	We performed a similar analysis for the fluid velocity at the particle position $\mathbf{u}(\mathbf{x_p},t)$ shown in \cite{SM} (Appendix F).
	Interestingly, the Einstein-Markov time-scale decreases with increasing $\textrm{St}$, unlike the behavior observed for $\mathbf{v}(t)$. 
	While this remarkable behavior and the effect this has on the fluid velocity observed by the particles requires further study, it may be related to the fact that a larger $\textrm{St}$ number implies a larger drag, and therefore a tendency of the system to dissipate energy more efficiently and faster. 
	This would result in a reduction of the associated Einstein-Markov time of 
$\mathbf{u}(\mathbf{x_p},t)$.

	The results shown above can be 
	linked to stochastic thermodynamics \cite{Nickelsen_2013, Reinke_2018} (also called stochastic energetics \cite{Sekimoto_2010}), developed for many different physical systems \cite{evans1993, cohen1995a, kurchan1998, Seifert_2005}, and to integral and detailed fluctuation theorems. 
	Experimental studies of stochastic thermodynamics mainly focus on nanoscale or quantum systems, or when dealing with classical systems, on biological systems \cite{collin2005, bechinger2015}, which are assumed to be well off the thermodynamic limit so that the probabilistic nature of balance relations becomes clearer.
	We test these theorems starting with the integral case as it should be more robust.
	
	For the case of passive particles, it is the turbulent surroundings that impose the non-equilibrium condition. Turbulence
	can be understood, and especially its energy cascade, as a process leading a fluid under specific conditions and parameters (the Reynolds number) from a non-equilibrium state into another non-equilibrium state, by the combination of energy injection and dissipation.
	In the spirit of non-equilibrium stochastic thermodynamics~\cite{Seifert_2012}, by using the Fokker-Planck equation, it is possible to associate with every individual trajectory $\left[u(\cdot)\right]$ a total entropy variation~\cite{Seifert_2005, Sekimoto_2010, Seifert_2012, Nickelsen_2013, Reinke_2018} given by the sum of two terms 
	\begin{eqnarray}
		\Delta S_{tot}\left[u(\cdot)\right] = \Delta S_{sys}\left[u(\cdot)\right] + \Delta S_{med}\left[u(\cdot)\right].
	\end{eqnarray}
	A thermodynamic interpretation of this quantity based on the relation between heat, work, and inner energy, can be given~\cite{Seifert_2005, Sekimoto_2010, Seifert_2012}. 
	$\Delta S_{sys}$ is the change in entropy associated with the change in state of the system for a single realization of a process that takes some initial probability $p\left(u_T, T\right)$ and changes to a different final probability $p\left(u_{\uptau_f}, \uptau_f\right)$.
	It is simply the logarithmic ratio of the probabilities of the stochastic process, so that its variation is
	\begin{eqnarray}
		\Delta S_{sys}\left[u(\cdot)\right]=
		-\ln{\left(
			\frac
			{p\left(u_{\uptau_f}, \uptau_f\right)}
			{p\left(u_T, T\right)} 
			\right)}.
	\end{eqnarray}
	The other term, the entropy exchanged with the surrounding medium $\Delta S_{med}$ from the initial to the final time scale, measures the irreversibility of the trajectories: 
	\begin{eqnarray}
		\Delta S_{med}\left[u(\cdot)\right]  
		&=& \int_{T}^{\uptau_f}
		\left[\partial_\tau u_\tau  \frac{D^{(1)} - \partial_{u_\tau}D^{(2)}/2 }{D^{(2)}}\right]d\tau.
		\label{eq:Entropy_med}
	\end{eqnarray}
	
\begin{figure}[ht]
		\includegraphics[width=0.237\textwidth]{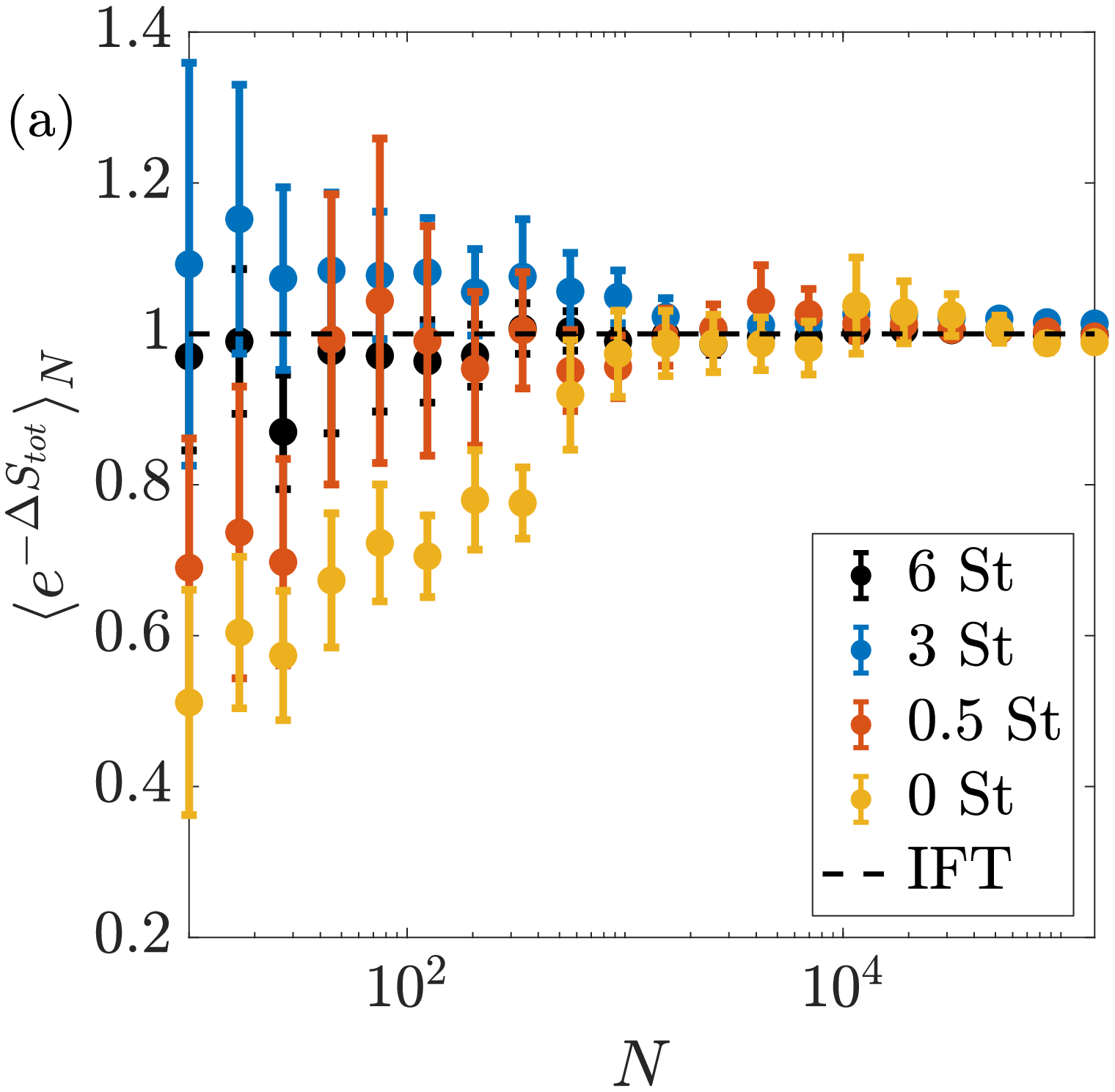}
		\includegraphics[width=0.22\textwidth]{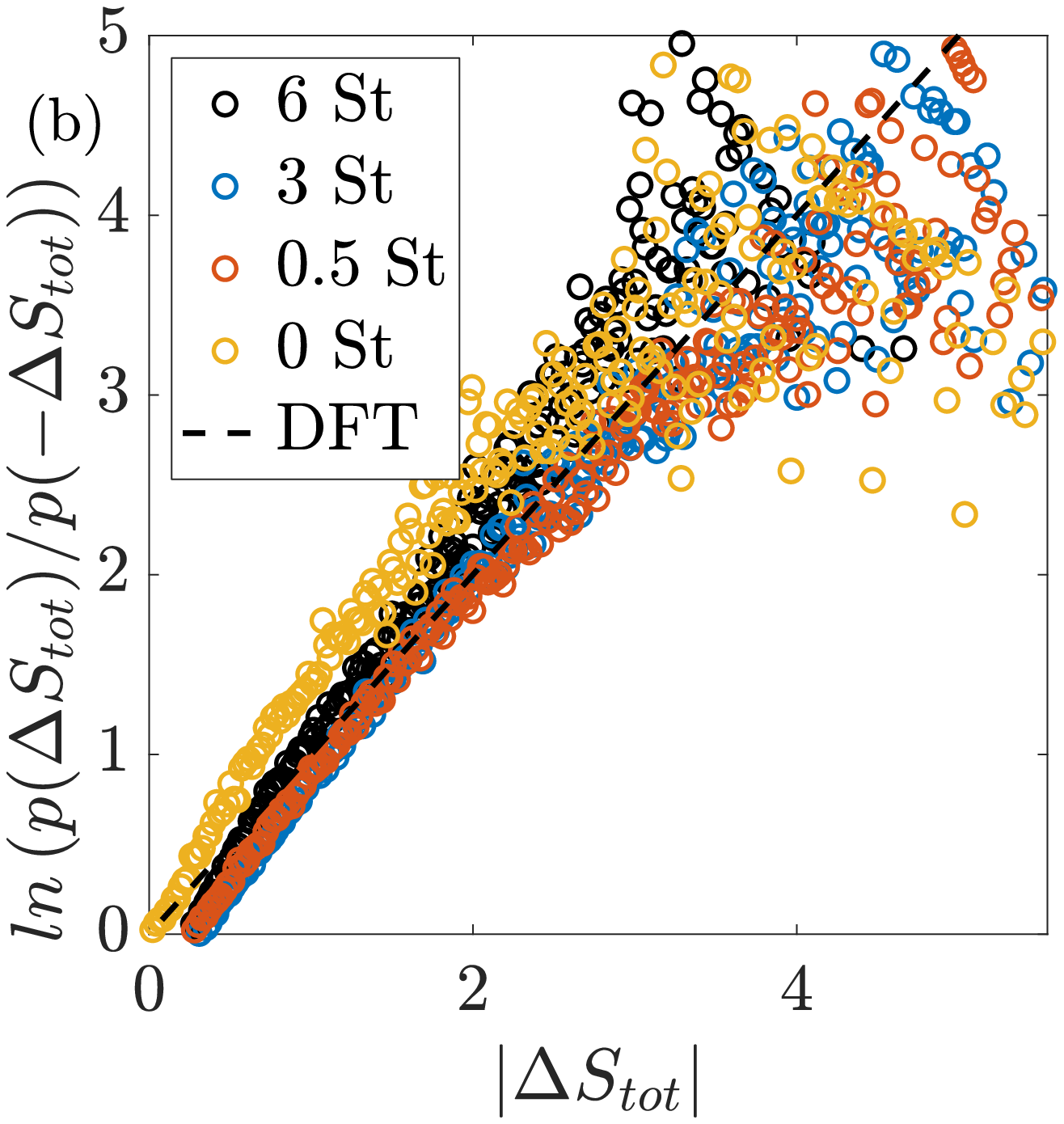}
		\caption{(a) Empirical average $\left\langle e^{-S_{tot}}\right\rangle_N=\frac{1}{N}\sum_{1}^{N}e^{-S_{tot}(N)}$ as a function of the sample size $N$ of $\left[u(\cdot)\right]$ trajectories. The set of measured cascade trajectories results in a set of total entropy variation values $\Delta S_{tot}$ (the same number of entropy values as the number of trajectories). According to the integral fluctuation theorem, the empirical average has to converge to a value of 1 (indicated by the horizontal dashed line). (b) Test of the detailed fluctuation theorem. The dashed line represents a linear behavior $p\left(\Delta S_{tot}\right) =  p\left(-\Delta S_{tot}\right) e^{\Delta S_{tot}}$.}
		\label{fig3}
	\end{figure}

	Then, the dynamics of 
	inertial particles coupled to a turbulent flow are characterized by the empirically estimated drift and diffusion coefficients $D^{(1,2)}$ (see \cite{SM} for a presentation of $D^{(1,2)}$ for
	all Stokes numbers considered).
	The knowledge of drift and diffusion coefficient allows to determine for each $\left[u(\cdot)\right]$ an entropy value, for which the thermodynamic fluctuations can be investigated.
	The dashed line in Fig.~\ref{fig3}(a) corresponds to the integral fluctuation theorem (IFT), which can be expressed as
	\begin{equation}
		\langle e^{-\Delta S_{tot}} \rangle_N = \int e^{-\Delta S_{tot}}p\left(\Delta S_{tot}\right) d\Delta S_{tot} = 1.
		\label{eq:IFT_stoch_theory}
	\end{equation}
	
	Figure \ref{fig3}(a) shows the empirical average 
	as a function of the number $N$ of trajectory sequences $\left[u(\cdot)\right]$ used for computing the average.
	The set of measured cascade trajectories results in a set of total entropy variation values $\Delta S_{tot}$ (the same number of entropy values as the number of trajectories). 
	We find that for the four Stokes numbers the results are in good agreement with the IFT, which is a fundamental entropy law of non-equilibrium stochastic thermodynamics \cite{Seifert_2005, Seifert_2012}. 
	This is not trivial, as the IFT is extremely sensitive to errors in the empirical estimation of $D^{(1,2)}$ defining the Fokker-Planck equation.
	For this reason, our empirical observation on the validity of the IFT for the inertial particles provides another verification of our approach.
	Furthermore, inertial particles (i.e., for sufficiently large $\textrm{St}$) sample preferentially some specific regions of the flow \cite{Mora_2021}. 
	The remarkable observation here is that they do it always in agreement with this theorem. 

	Figure \ref{fig3}(b) shows that in addition to the IFT also the detailed fluctuation theorem (DFT) holds for our data coming from a turbulent flow with finite Reynols number, which is expressed as
	\begin{eqnarray}
		\label{eq:DFT}
		\ln{\left(\frac{p\left(\Delta S_{tot}\right)}{p\left(-\Delta S_{tot}\right)}\right)} &=& \Delta S_{tot}.
	\end{eqnarray}
	Thus, in addition to the IFT, the DFT expresses the balance (explicit exponential symmetry constraint) between entropy-consuming ($\Delta S_{tot}<0$) and entropy-producing ($\Delta S_{tot}>0$) trajectories.
	See \cite{SM} for the illustration of $p\left(\Delta S_{tot}\right)$ for \mbox{$\textrm{St}=0.5$, $3$ and $6$}.
	
\begin{figure}[ht]
		\includegraphics[width=0.23\textwidth]{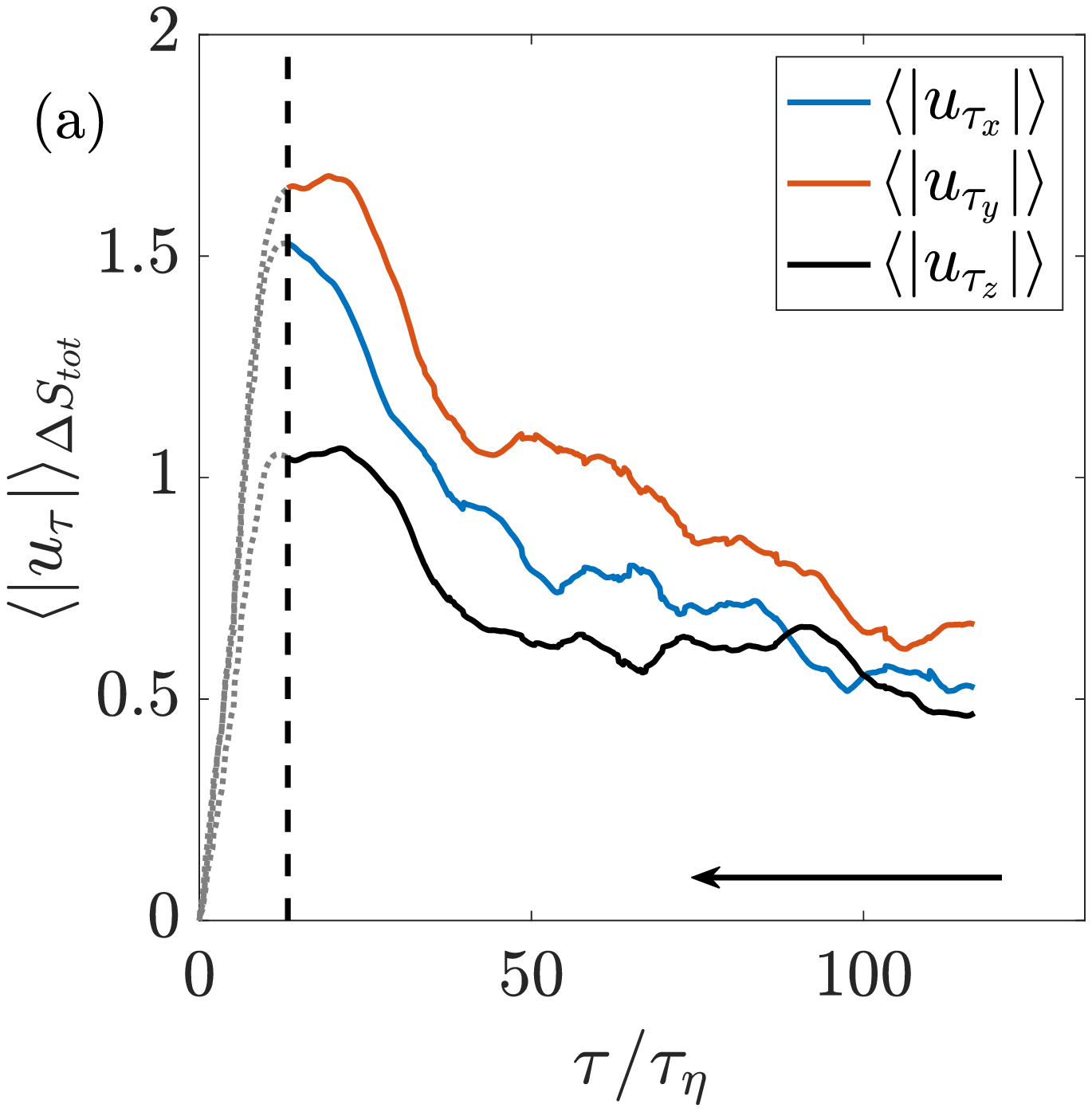}
		\includegraphics[width=0.23\textwidth]{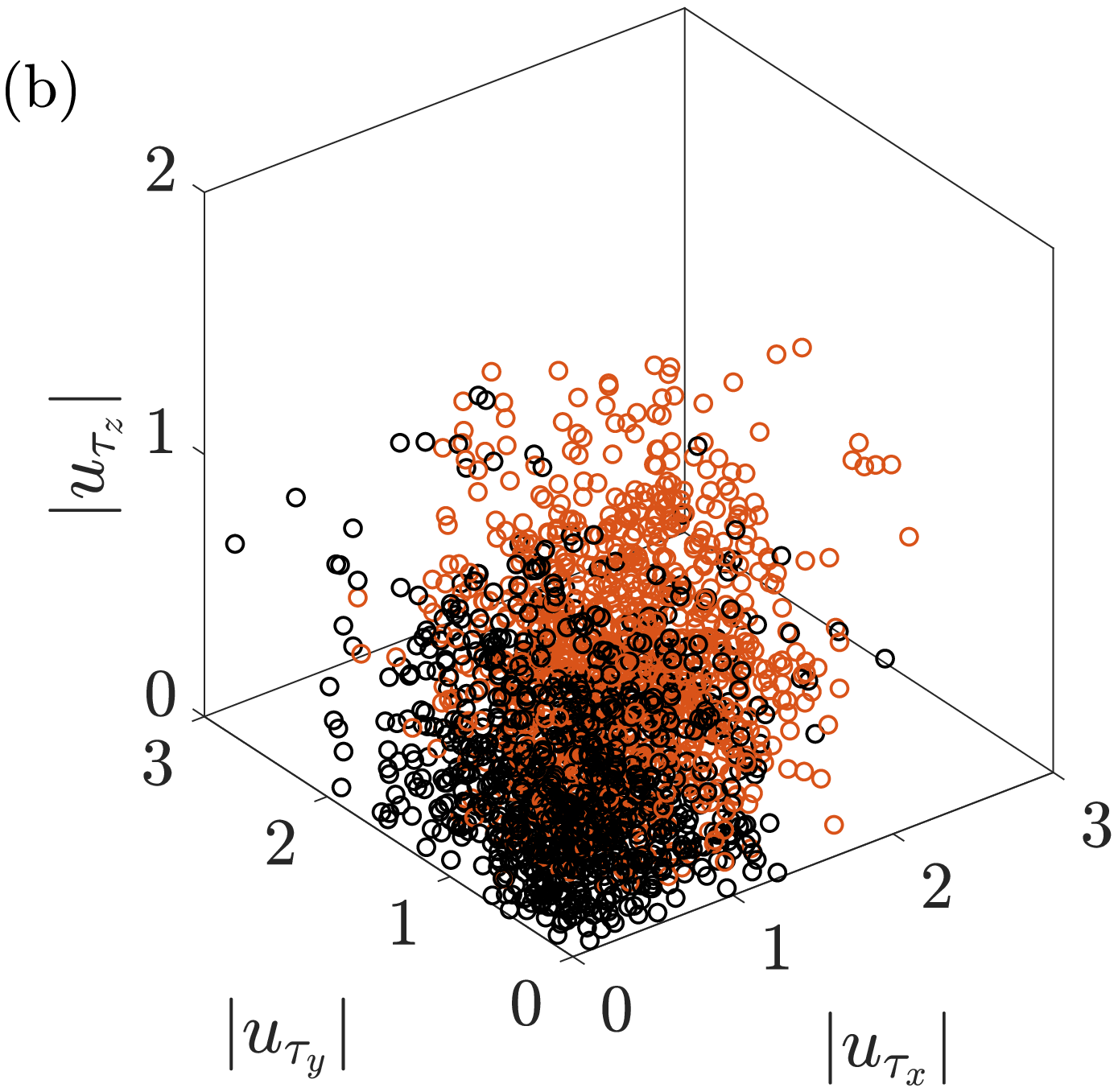}\\
		\includegraphics[width=0.23\textwidth]{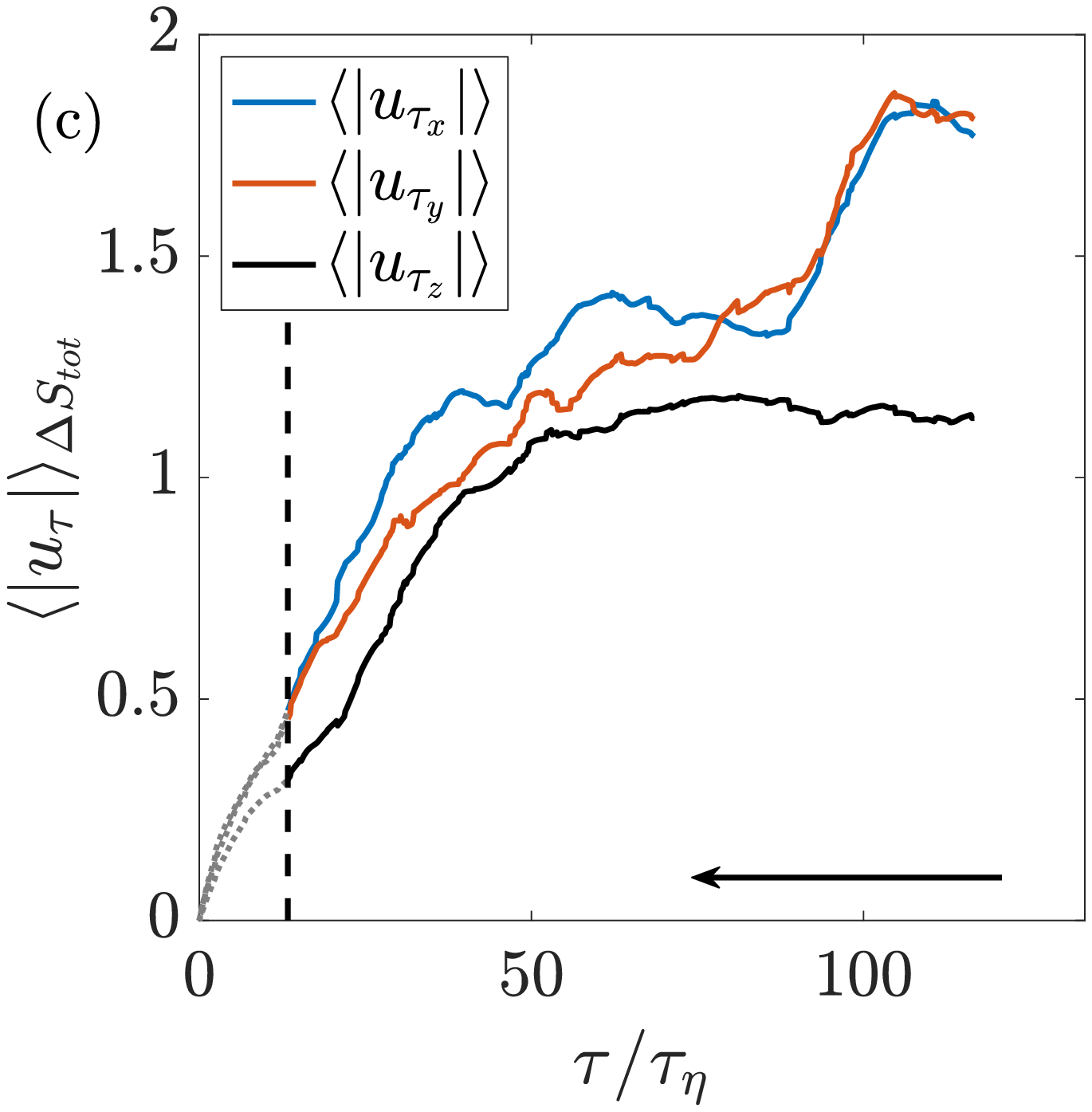}
		\includegraphics[width=0.23\textwidth]{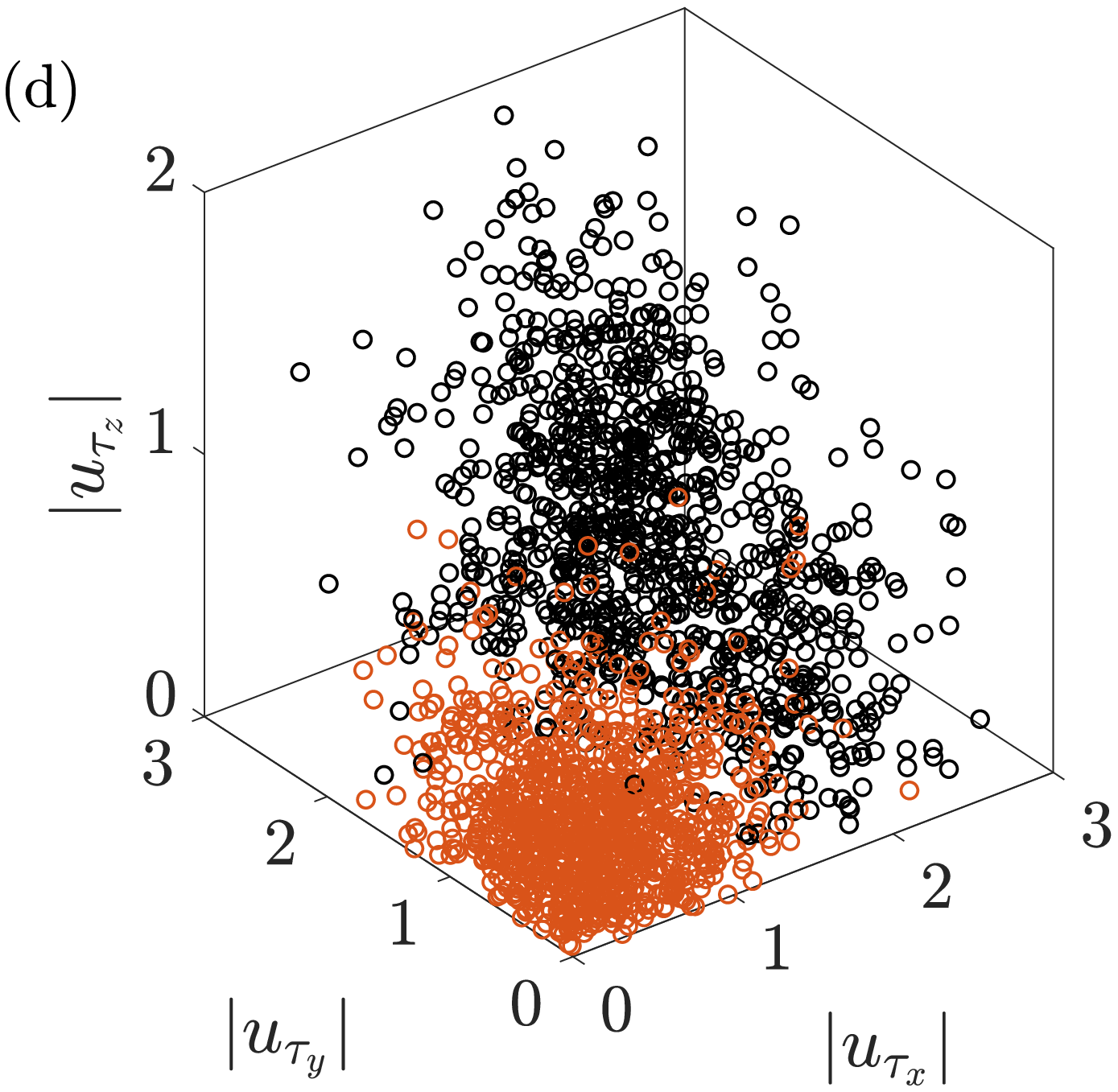}
		\caption{Mean absolute velocity increment trajectories conditioned on a specific total entropy variation $\langle |u_{\tau}| \rangle_{\Delta S_{tot}}$, \mbox{$\Delta S_{tot}=-2$ (top)} and $\Delta S_{tot}=3$ (bottom). Left: individual representation of the mean absolute components for $x$, $y$ and $z$. The vertical dashed line marks the Einstein-Markov coherence time scale $\Delta_{EM}$. The dissipation region is indicated by dotted lines. The arrow indicates the direction of the path through the hierarchy of time scales $\tau$, from initial time scale $T$ to the final time scale $\uptau_f$. Right: Circles mark increment on the initial (black) and $\Delta_{EM}$ (red) scale in a three dimensional scatter plot. In these scatter plots increments of all trajectories for which the conditional averaging was performed are shown.}  
		\label{fig4}
	\end{figure}

	Finally we ask whether these findings of the entropy fluctuation are just a statistical result or if they have some relation to flow structures. 
	In particular, it is of interest if the negative entropy events have some special features.
	Therefore we study the mean absolute velocity increment trajectories conditioned on a specific total entropy variation $\langle |u_{\tau}| \rangle_{\Delta S_{tot}}$. 
	Figures \ref{fig4}(a) and (c) show that the total entropy variation is linked to distinct trajectories. 
	Entropy-consuming trajectories are characterized by an increase in the averaged absolute values of the increments $\langle |u_{\tau}| \rangle_{\Delta S_{tot}=-2}$ with decreasing time scale $\tau$, while trajectories marked by entropy-production $\langle |u_{\tau}| \rangle_{\Delta S_{tot}=3}$ smoothly decrease their absolute increments with decreasing $\tau$. 
	This is further highlighted by Fig. \ref{fig4}(b) and (d). 
	In these three-dimensional representations the absolute velocity increment at initial and final scale, $T$ and $\Delta_{EM}$, conditioned on the entropy for all trajectories for which the conditional averaging was performed, are shown.
	In \cite{SM} (Appendix E) we show that the majority of large and intermittent small scale increments (heavy tailed statistics of the small-scale increment PDF) are contained in the negative entropy trajectories.\\

	\section{Conclusion}
	In this letter we showed the existence of Markov property for Lagrangian turbulence.  
	The dynamics of inertial particles coupled to a turbulent flow 
	can be treated as a Markov process for a finite step size larger than a Stokes number-dependent Einstein-Markov coherence time scale. 
	This opens an interesting interpretation of the $\textrm{St}$ number in terms of the Markov memory of the particles trajectories, and is compatible with the picture that particles with more inertia filter fast and small scale fluctuations of the carrying flow: for particles with larger $\textrm{St}$ (and thus larger particle response times) the Markovianization of trajectories by the turbulence takes place at longer times. 
	This can be used to quantify effective parameters in cases where the particles' inertia is not clearly defined, like finite-size, or non-spherical particles.
	
	Moreover, based on the theory of Markov processes
	it is possible to derive an explicit Fokker-Planck equation for the description of multi-scale statistics of Lagrangian particles.
	The Markov property corresponds to a three-point (two-scale) closure of the general joint multi-scale statistics \cite{Peinke2018}, representing a step towards a more comprehensive understanding of turbulence, which is characterized by a challenging complexity with many high order correlations. 
	This results in a tremendous reduction of the degrees of freedom, and shows that the long-standing problem in turbulence of the closure of the hierarchy of equations can be approached in a Lagrangian framework.

	The interpretation of the stochastic process 
	as an analogue of a non-equilibrium thermodynamic process \cite{Nickelsen_2013,Reinke_2018,Peinke2018}
	when applied to Lagrangian turbulence allowed us to show that inertial particles satisfy both integral and detailed fluctuation theorems in a strict sense.
	The verification of these theorems, which have contributed significantly to the understanding of open and out-of-equilibrium systems and to the development of statistical physics, adds strength to the validity of the Markov property and the description by a Fokker-Planck equation.
	Previous observations of the irreversibility of particles in turbulence considered a Jarzynski-like equality for the particles energetics \cite{Xu_2014}, but the slope was not one as expected for the DFT. 
	Reproductions of such exact results 
	are found very rarely in turbulence research, so that the result provides a novel constraint for the particles' evolution.  
	The fluctuation theorems may be taken also as a kind of new conservations laws.

	These fluctuation theorems also shed light on asymmetries in particle trajectories, as reported before in \cite{Xu_2014}, and as shown here for individual trajectories depending on their entropy evolution. 
	A connection can be made between entropy-producing and -consuming trajectories with specific evolution of velocity increments 
	with decreasing time scale.
	Interestingly, our study also shows that particle trajectories are out-of-equilibrium while keeping ergodicity. 
	These abnormal behaviors (or structures) in the Lagrangian paths are statistically embedded in the fulfillment of the two fluctuation theorems.
	Thus, the presented results may shed new light on the long-lasting debate of the statistical approach to turbulence versus the study of flow structures.

	The case of out-of-equilibrium systems like turbulence coupled to dense particles
	has applications in geophysics, astrophysics, and aerosol dispersion.
	In the cases studied here we have non-trivially coupled systems, whose particles not always randomly sample the flow topology \cite{Mora_2021}. 
	Note that preferential concentration for $St \neq 0$ implies that particles are not sampling the flow homogeneously, and a non-random sub-sampling of the Eulerian flow velocity, plus the particles own inertia, could break down the IFT and DFT theorems. 
	In spite of this, the theorems are satisfied for particles with different response times (albeit, in the limit $\textrm{St} \to \infty$, the relations should not hold as particles decouple from the flow). 
	Also,
	the formalism proposed here only requires measuring the trajectory and velocity of particles. 
	Such study is accessible with state-of-the-art experimental techniques, even at large Reynolds numbers, while most models to describe the dynamics 
	of these systems require much more detail, such as measurements also of particles’ acceleration.
	At last, we point out that the entropy analysis may also serve to compare different turbulent situations, as it was successfully done by the statistics $p(\Delta S_{tot})$ for turbulent wave situations \cite{hadjihoseini2018rogue}.

	\acknowledgments
	This work has been partially supported by the ECOS project A18ST04, by the Volkswagen Foundation (96528) and by the Laboratoire d’Excellence LANEF in Grenoble (ANR-10- LABX-51-01).
	
\newpage
	\bibliographystyle{eplbib}
	\bibliography{biblio}
\end{document}